\documentstyle[12pt]{article}
\begin{document}
\overfullrule=0pt

\addtolength{\baselineskip}{2mm}
\addtolength{\parskip}{1mm}

\newcommand{\by}{\bar y}
\newcommand{\ah}{{1 \over 2}}
\newcommand{\ra}{\rangle}
\newcommand{\la}{\langle}

\newcommand{\mf}{I}
\newcommand{\ms}{I \hskip -0.5mm I}
\newcommand{\mt}{I \hskip -0.5mm I \hskip -0.5mm I}
\newcommand{\tg}{\tilde g}
\newcommand{\te}{\tilde e}
\newcommand{\tjr}{\tilde j^q}

\newcommand{\bya}{\bar y_1}
\newcommand{\byb}{\bar y_2}
\newcommand{\tm}{\tilde G^{\Delta}}

\begin{titlepage}
\begin{center}

{\Large \bf  Bimodule Properties of Group-Valued Local Fields}
{\Large \bf  and Quantum-Group Difference Equations}

\vspace{3cm}

{\bf Ling-Lie Chau and Itaru Yamanaka} \\[1mm]

\vspace{5mm}

{\it Physics Department, University of California, Davis, CA 95616}
\end{center}

\vfill
\abstract{
We give an explicit construction of the quantum-group generators ---
local, semi-local, and  global --- in terms of the group-valued  quantum
fields
$\tilde g$ and $\tilde g^{-1}$  in the Wess-Zumino-Novikov-Witten (WZNW)
theory. The algebras among the generators and the fields make concrete and
clear the
bimodule properties  of the $\tilde g$ and the $\tilde g^{-1}$ fields.
 We show that  the correlation
functions of the $\tilde g$ and $\tilde g^{-1}$ fields in  the vacuum state
defined through the semi-local quantum-group generator satisfy a set of
quantum-group
difference equations. We give the explicit  solution for the two point
function.
A similar formulation can also be done for the quantum Self-dual Yang-Mills
(SDYM)
theory in four dimensions.
}

\vfill
\end{titlepage}

The Wess-Zumino-Novikov-Witten (WZNW)  theory \cite{wz,w} has a long and
beautiful history.  In his well known 1984 paper  Witten \cite{w}
quantized the Lie-algebra valued field $ \tilde j_\mu$ of the theory and
derived its current algebra with central
charge.  From this current algebra  Knizhnik and Zamolodchikov \cite{kz}
derived
the linear equations  (the K-Z equations) satisfied by the correlation
functions. (In their original formulation the
Virasoro generators \cite{v} played an important role. Actually one can
obtain
the K-Z equations without involving the Virasoro-algebra generators
 \cite{fr}.)  Later the quantum-group  structures of the
theory were discovered \cite{fa1,dj} and studied
 in many papers,  \cite{tk} to \cite{mr}.
 However,
in
all of these studies the role of the group-valued local quantum fields
$\tilde g$,
the basic fields  in the theory, was not clear.

Recently the WZNW theory was studied from the point of
view of considering  the fields $\tilde g$ as
the basic fields, papers \cite{fa2} to \cite{cy2}.   We quantized  the
group-valued local
quantum
fields $\tilde g$ of the Wess-Zumino action in the light-cone
coordinates \cite{cy1,cy2}
 using the Dirac procedure for constrained systems \cite{d}.  Further, we
had
also  formulated the quantum
Self-dual Yang-Mills (SDYM) system \cite{cy3} in terms of the group-valued
local
field
$\tilde J $ and showed how  the two
theories are related. The
quantum WZNW theory in terms of
$\tilde g$ can be obtained from the quantum SDYM theory in terms of
$\tilde J$  by reducing the two appropriate dimensions in the
quantum SDYM theory.   The exchange
algebras satisfied by the group-valued local quantum fields in the two
theories are
very similar.  In both cases we showed that the second-class
constraints in forming the Dirac brackets in the light-cone coordinates are
the source
of  the nontrivial critical exponents in the products
of fields and the quantum-group structures in these theories. However there
are very
important differences between the two theories. The WZNW theory is a free
theory in
the light-cone-coordinate formulation in two dimensions.  The self-dual
Yang-Mills
theory is an interacting theory (even in
the light-cone coordinate formulation) in
four
dimensions. One can easily see how the interactions disappear in the
dimension
reduction. Because of its simpler structure,  the quantum WZNW theory is an
important
laboratory for the study of the quantum SDYM field theory, which in turn is
an
important laboratory for the study of many other quantum field theories in
four
dimensions \cite{c1}.

 In addition to the difference in our way of obtaining the exchange algebra
from
those of Refs.~\cite{fa2} to \cite{cghos}, there are other important
differences in our
interpretation of the exchange algebra and in the further development of
theory.
We have given an analytic interpretation to the spatial dependence of the
$R$
matrix of the exchange algebras of the $\tilde g$ fields. From that
interpretation we
have formulated the normal-ordering procedure, constructed the $\tilde
g^{-1}$ quantum
fields and their exchange algebras, constructed the
Lie-algebraic
current $\tilde j(\bar y)
\sim
\tilde g\partial_{\bar y} \tilde g^{-1},$ and derived the current
algebra from the exchange algebras of $\tilde g$ and
$\tilde g^{-1}$ without resorting to the use of the
boson quantum fields \cite{com1}. This procedure also makes it
straightforward to
construct  the theory for the general sl(n) cases.

     What has  emerged is that the group-valued local quantum
fields, $\tilde g$  and $\tilde g^{-1}$
 are bimodule quantum fields.
 Dictated by the $R$ matrix and the nontrivial
 critical exponents, the $\tilde g$ fields form noncommutative vector
spaces of the quantum-group on the right side and commutative vector spaces
on the left side.
The left side of $\tilde g$ form the fundamental representation of the Lie
group and the Lie-algebra currents $\tilde j$ are the generators for its
transformation. The right side of $\tilde g$ forms the representation
of the quantum-group.
 However, until now the
generators of the quantum-group transformations had not been fully
constructed
and it was unknown whether or not  they could be constructed from
$\tilde g$ and $\tilde g^{-1}$ quantum fields.
 (The above statements
apply similarly
to   $\tilde g^{-1}$, except that the roles of the two sides are
reversed.)

In this paper we give an explicit construction of the quantum-group
generators ---
local, semi-local, and  global ones --- in terms of the quantum $\tilde g$
and $\tilde g^{-1}$  fields. Their algebras make clear and concrete the
bimodule
properties  of the $\tilde g$ and the $\tilde g^{-1}$ fields. From the
semi-local quantum-group generators, we define a vacuum which we call the
$U_q^{\Delta}[\hbox{sl}(n)]$-vacuum.  It is different
from the
vacuum defined from the regular Lie-algebra current  $\tilde j(\bar y)$,
which we call
the $\widehat
{\hbox{sl}(n)}$-vacuum. The two vacua are different
 since the semi-local quantum-group generators do not
commute
with the Lie-algebra current.  We then show that the
$U_q^{\Delta}[\hbox{sl}(n)]$-vacuum-expectation-values of
products of the
$\tilde g$ and $\tilde g^{-1}$ fields satisfy a set of linear difference
equations, which we call the quantum-group difference equations.
(These  equations look different from
those discussed in Ref.~\cite{fr}. )   For
the
two point function we provide the solution.
With these additional understandings,
the bimodule properties of the WZNW theory become concrete and clear.

The bimodule properties, we believe, are generic for all group-valued
local quantum   fields  as we  showed earlier \cite{cy3}
 that they also
 hold in the
SDYM
quantum field theory.  For the interacting SDYM quantum field
theory the
exchange algebras of the  group-valued
quantum local fields $\tilde J$ and $\tilde J^{-1}$are just the starting
point of the
theory.  The exchange algebras and the
current algebras in the SDYM theory are
fixed-time relations to begin with. Using one of the
additional
dimensions available and performing its
spatial-integration,  in paper~\cite{cy3} we  constructed time-independent
currents and their current algebras, with interesting features of
higher dimensions.   From these
time-independent
 currents
\cite{rtt}
we are able
to
construct a time-independent local monodromy matrix  $\tilde T \circ$ and
derived its local exchange algebra
$R^T \tilde T \circ \tilde T \circ =
\tilde T \circ  \tilde T \circ R^T$ which contains the infinite local charge
algebras and the infinite local Yangian algebras
 and then derive
differential
and difference equations for their correlation functions in the SDYM
theory \cite{cy4}.

The organization of the rest of the paper is the following. We first give a
brief
description of our formulation of the quantum WZNW theory.
To be  specific and
simple, we discuss here the case of  $\hbox{sl}(2)$. Our formulation  can be
generalized to the cases of $\hbox{sl}(n)$ in a straightforward
way  \cite{com2}.
 We then give the
construction of the
quantum-group generators and  their algebras.  After defining the
$U_q^{\Delta}[\hbox{sl}(2)]$-vacuum through the semi-local quantum-group
generators
$ \tilde G^\Delta(\bar y)\equiv \tg^{-1}(\bar y-\Delta)~\tg(\bar y +\Delta)$,
we show that the correlation functions of  the  $\tilde g$ and $\tilde
g^{-1}$ fields defined in this vacuum satisfy a
set of  difference equations.   For the two point function we give the
explicit
solution. We then end the paper with some concluding remarks.

\smallskip

{\bf \it Exchange Algebras, Critical Exponents, Normal-Ordering, and
Current Algebras }

       In this section we briefly describe the essential points of our
formulation of the quantum WZNW theory in terms of the  $\tilde g$ and
$\tilde
g^{-1}$ fields so that the new development can be clearly presented.
 In the case of $\hbox{sl}(2)$, $\tilde g$ is a
$2\times 2$ matrix with non-commuting operator-valued entries, which we call
the quantum $\tilde g$ fields. They satisfy the following exchange algebras,
\begin{equation}
 \tilde g_\mf(\bya)\tilde g_{\ms}(\byb) =
1_{\mf,\ms}\tilde g_{\ms}(\byb)\tilde g_\mf(\bya)
R_{\mf,\ms}(\bya-\byb), \label{1}
\end{equation} where

\begin{equation}
R_{\mf,\ms} =
P_{\mf,\ms}\left\{[q]^{\triangle{_1}\ \varepsilon(\bya-\byb)}
{\cal P}_{j_{12} = 1}^q-[q]^{\triangle{_0}\ \varepsilon(\bya-\byb)}
{\cal P}^q_{j_{12}=0}\right\}.   \label{2}
\end{equation}

\noindent Here we denote the light-cone coordinate $x-t\equiv \bar
y$. The time in this light-cone-coordinate formulation is $y \equiv x+t$. In
Eq.~(\ref{2}),
\begin{equation}
q\equiv
e^{-i\hbar/4\alpha},
\end{equation}
 which becomes unity when $\hbar
\rightarrow 0$;
$\alpha
$ is the coefficient in front  of the Wess-Zumino action;
$\triangle{_1}=2\ah (\ah+1)-1(1+1)=-1/2$ and
$\triangle{_0}=2\ah (\ah+1)-0(0+1)=3/2$
are the differences of conformal dimensions of two spin ${1 \over 2}$
fields minus that of a spin $j_{12}=1$ and $0$ fields, respectively
;  the subscripts I and II denote the
tensor spaces that
the
operator  matrices or c-number matrices operate on. (This tensor notation
saves us
the
trouble of writing out the indices of the matrix elements.
Written in terms of the matrix elements, Eq.~(\ref{1}) reads
$ \tilde g_{m_1,\alpha_1}(\bya)\tilde g_{m_2,\alpha_2}(\byb) =
\delta_{m_1,l_1}\delta_{m_2,l_2}
\tilde g_{l_2,\beta_2}(\byb)\tilde g_{l_1,\beta_1}(\bya)
R_{\beta_1,\beta_2;\alpha_1,\alpha_2}(\bya-\byb)$, where the repeated indices
$l_1, l_2, \beta_1$,
 and
$\beta_2$ are summed.) The ${\cal P}^q_{j_{12}}$'s
are the $4\times 4$ c-number $q$-ed projection matrices projecting
the two spin 1/2 states into $j_{12}=0$ or 1,
 satisfying ${\cal P}^q_{j_{12}}
{\cal P}^q_{j_{12}'} = {\cal P}
^q_{j_{12}} \delta_{j_{12}j_{12}'}$,

\begin{equation}
{\cal P}^q_{j_{12}=1} = {\rm diag} \{1,a\left({q \atop 1}{1
\atop q^{-1}}\right), 1\}, \quad {\cal P}^q_{j_{12}=0}=1-
{\cal P}^q_{j_{12}=1},
\label{2.a}
\end{equation}
%
%$$\eqalign{
%{\cal P}^q_{j_{12}=1} &= {\rm diag} \{1,a\left({q \atop 1}{1
%\atop q^{-1}}\right), 1\}, \cr
%}\eqno(2.a)$$
\noindent where $a\equiv 1/(q+q^{-1})\equiv 1/[2]_q$, with $[n]_q\equiv
(q^n-q^{-n})/(q-q^{-1})$.  The matrix $P_{\mf,\ms}$
 interchanges matrices in space I to II and visa versa, e.g.,
$P_{\mf,\ms}~\tilde g_\mf(\bya)~\tilde g_{\ms}(\byb)=
\tilde g_{\ms}(\bya)~\tilde g_\mf(\byb)~P_{\mf,\ms}$, and its explicit
representation is $P_{\mf,\ms}
={1\over 2}(1-\sum^3_{a=1}\sigma^a\sigma^a)={\cal
P}_{j_{12}=1} - {\cal P}_{j_{12}=0}$; here the ${\cal P}_{j_{12}}$'s are the
un-$q$-ed ordinary projection matrices, i.e., Eq.~(\ref{2.a}) with $q=1$.

The $q$-deformation in the $R$ matrix, i.e., its quantum-group structure, is
an
$\hbar$ effect, $R=1$ when $\hbar=0$. Taking the leading term in $\hbar$,
Eq.~(\ref{1})
becomes the Dirac bracket of the $g$ fields. Therefore we directly identified
that
the second-class constraints are the source of the quantum-group structure of
the
theory, \cite{cy1} and \cite {cy2}.

The $\varepsilon(\bya-\byb)$ in Eq.~(\ref{2}) is a signature of the
quantizaton in the light-cone coordinates.
It has the usual definition
\begin{equation}
\varepsilon(\bya-\byb)=\pm1,~ \quad{\rm for~~} \bya ~{}^{>}_{<}
{}~\byb.~~~  \label{2.b}\end{equation}
We denote
$R_{\mf, \ms}(\bya-\byb)=R_{\mf, \ms}(+)$, for $\bya-\byb>0$ and
$R_{\mf, \ms} (\bya-\byb)=R_{\mf, \ms}(-)$, for $\bya-\byb<0$.  Note
that $[R_{\mf, \ms}(+)]^{-1} =R_{\ms, \mf}(-)$ and $[R_{\mf,
\ms}(-)]^{-1}=R_{\ms, \mf}(+)$.
In our formulation we emphasize the analytic-function interpretation of
$\varepsilon(\bya-\byb)$, i.e.,
\begin{equation}
\varepsilon(\bya-\byb)
=-[ln(\bya-\byb+i\varepsilon)-ln(\byb-\bya+i\varepsilon)]/\pi i,
\label{2.c}\end{equation}
and define
\begin{equation}
\varepsilon(\bya-\byb)=0,~  \quad{\rm for~~} \bya = \byb.
\label{2.d}\end{equation}
The expression for $\epsilon(\bya-\byb)$, Eq.~(\ref{2.c}), indicates that the
product $\tilde g_\mf (\bya) \tilde g_{\ms}(\byb)$ has singularity at
$\bya-\byb=0$ with  critical exponents given by
\begin{equation}
{\cal P}_{j_{12}} \tilde g_\mf(\bya) \tilde g_{\ms}(\byb)
{\cal P}^q_{j'_{12}}=(\bya-\byb)^{ \triangle_{j_{12}}(ln \
q)/\pi i}\{:{\cal P}_{j_{12}}\tilde g_\mf(\bya)\tilde
g_{\ms}(\byb){\cal P}^q_{j'_{12}}:\}.\label{3}\end{equation}
This also defines the normal-order products to be those in the curly
brackets;  their Taylor expansions give the operator-product expansions.

Using Eq.~(\ref{2.d}), we can
easily prove that at $\bya=\byb$, the exchange algebra Eq.~(\ref{1}) gives
\begin{equation}
{\cal P}_{j_{12}}g_\mf(\bya)g_{\ms}(\bya)=g_\mf(\bya)
g_{\ms}(\bya){\cal P}^q_{j_{12}}, \label{2.e}\end{equation}
where $ j_{12}=0 , 1$. Eq.~(\ref{2.e})  implies ${\cal P}_{j_{12}}g_\mf(\bya)
g_{\ms}(\bya)
 {\cal P}^q_{j'_{12}}=0$, for $j_{12}\not=j'_{12}$. This fact and the later
development of the quantum-group generators rely crucially on the
interpretation of
the $R$ matrix at the coincidence point, Eq.~(\ref{2.d}).

The quantum  $\tilde g$ field can be interpreted as a non-Abelian vertex
operator.
 However, notice that our $\tilde g$ field in component form has the bimodule
indices. The vertex
operators given in Refs.~\cite{msp}, ~\cite{ags}, and ~\cite{mr} have one
less index and no bimodule
property indicated.

We defined the $\tilde g^{-1}$ field from the following equation
\begin{equation}
\tilde g(\bar y) \tilde g^{-1}(\bar y)=1
= \tilde g^{-1}(\bar y)\tilde g(\bar y).\label{4}\end{equation}

\noindent From Eqs.~(\ref{4})~and~(\ref{1}), we can easily show that the
$\tilde
g^{-1}$ field
satisfies the following exchange algebras
\begin{equation}
\tilde g^{-1}_\mf(\bya) \tilde g_{\ms}(\byb) = \tilde
g_{\ms}(\byb) [R_{\mf,\ms}(\bya-\byb)]^{-1} \tilde
g^{-1}_\mf(\bya),\label{5a}\end{equation}
\noindent and
\begin{equation}
\tilde g^{-1}_\mf(\bya) \tilde g^{-1}_{\ms}(\byb) =
R_{\mf,\ms}(\bya-\byb) \tilde g^{-1}_{\ms}(\byb) \tilde
g^{-1}_\mf(\bya). \label{5b} \end{equation}
\noindent The construction of  this $\tilde g^{-1}$ field is  crucial for us
to
develop the full content of the theory in terms of the group-valued fields.

{}From $\tilde g$ and
$\tilde g^{-1}$, we constructed
 the  $\widehat {\hbox{sl}(2)}$ current
\begin{equation}
\tilde j(\bar y) \equiv 2\alpha  ~ \tilde g\partial_{\bar y} \tilde
g^{-1}. \label{6}\end{equation}
 We then showed that the following equations can
be easily derived from
the exchange algebras of the $\tilde g$
and $\tilde g^{-1}$  quantum fields,
\begin{eqnarray}
{[ \tilde j_{\mf}(\bya), \tilde j_{\ms}(\byb) ]} &=&  -i\hbar [M_{\mf,\ms},
\tilde j_{\ms}(\bya)]
{}~\delta (\bya-\byb) -i\hbar 2 \alpha M_{\mf,\ms}~\delta'(\bya-\byb),
\label{7}
\\
{[\tilde j_{\mf}(\bya), \tilde g_{\ms}(\byb)]} &=&  -i\hbar M_{\mf,\ms}~
\tilde g_{\ms}(\bya)   \delta (\bya-\byb),
\label{8} \\
{[\tilde j_{\mf}(\bya), \tilde g^{-1}_{\ms}(\byb)]}  &=&  i\hbar  \tilde
g^{-1}_{\ms}(\bya)
M_{\mf,\ms}~   \delta (\bya-\byb),
\label{9}
\end{eqnarray}
where $M_{\mf,\ms}\equiv P_{\mf,\ms}-{1\over 2}={1\over
2}\sum^3_{a=1}\sigma^a_\mf\sigma^a_{\ms}$. Eq.~(\ref{7}) is the current
algebra
of the current $\tilde j$. Taking trace of
 Eq.~(\ref{7}) onto ${1 \over 2i}\sigma_\mf^a$ and ${1 \over
2i}\sigma_{\ms}^b$ one can easily obtain
 the more familiar form  of the current algebra  in terms of the
 Lie-components of the current $[\tilde j^a(\bya),\tilde
j^b(\byb)]=i\hbar \varepsilon^
 {abc}\tilde j^c(\bya)
{}~\delta(\bya-\byb)  -
{i \hbar^2 \over 4 \pi} K
 \delta^{ab} ~\delta'(\bya-\byb)$, where $K \equiv i\pi [ln(q)]^{-1}
= - 4\pi \alpha / \hbar$.
Therefore we had reproduced the well known current algebra given by Witten in
Ref.~\cite{w},
but we have constructed from the group-valued quantum fields $\tg$ and
$\tg^{-1}$, a different
quantum formulation of WZNW theory. Eq.~(\ref{8})
indicates that the left side of $\tilde g$ forms the
fundamental representation of the
 current $\tilde j$; Eq.~(\ref{9}) indicates that the right side of
$\tilde g^{-1}$
 forms the fundamental representation of the current $\tilde j$.
 From
\begin{equation}
 2\pi i~\delta(\bya-\byb)={1\over
 \bya-\byb-i\varepsilon}-{1\over \bya-\byb+i\varepsilon},
\end{equation}
the
$\delta$-function
 on the right-hand-side of Eqs.~(\ref{7}) to (\ref{9}) indicats that those
products
 of fields have singularities.
 Equations Eqs.~(\ref{7}) to (\ref{9}) can be equivalently written out as
 operator-product-expansions for products of operators (which we leave as
exercises
for the reader).  Next we present our new development.

\smallskip

{\bf \it Quantum-Group Currents and Global Quantum-Group Generators  }

 Similar to the construction of the current $\tilde j(\bar y)$, it is nature
to construct the other current $ \tjr(\bar y)$

\begin{equation}
 \tjr(\bar y) \equiv 2\alpha ~\tg^{-1}(\bar y) \partial_{\bar y}
\tg(\bar y), \label{10}\end{equation}
which we shall call
the quantum-group current
since it has the quantum-group index on both sides. We can work out the
algebraic relations among the matrix elements of
$\tjr(\bar y)$, corresponding to Eq.~(\ref{7}) for $\tilde j(\bar y)$; and
the
algebraic
relations with the fields
$\tilde g$ and $\tilde g^{-1}$, corresponding to Eqs.~(\ref{8}) and
(\ref{9}). All of
them  have
nice quantum-group interpretations.  However, we find that $\tjr$ is not as
useful a
quantity as the current $\tilde j$ in that it can not be used to develop its
vacuum
states and the corresponding differential equations as the current
$\tilde j$ was used to develop the K-Z equations. Therefore here we do not
write out
those algebraic relations involving $\tjr$.

On the other hand, we find that the following group-valued quantities,
$\tilde G$
and $\tilde G^{\Delta}$, are the more useful quantum-group generators.
We form the global quantum-group generator  $\tilde G$  from  the
quantum-group current
$\tjr$ of Eq.~(\ref{10}) by a path ordered
 integration,
\begin{equation}
 \tilde G = \vec P exp( \int_{-\infty}^{\infty}d{\bar
y}~\tg^{-1}\partial_{\bar y}
\tg)=\tg^{-1}(-\infty)\tg(\infty)~~.\label{11} \end{equation}

{}From the exchange algebras, Eqs.~(\ref{1}), (\ref{5a}), and (\ref{5b}), we
can derive the
algebraic relations between the matrix elements of $\tilde G$ and the action
of
$\tilde G$ on $\tg$ and $\tg^{-1}$,
\begin{eqnarray}
\{ R_{\ms , \mf}(+)~ \tilde G_\mf~ R_{\mf , \ms}(+)~ \} \tilde G_{\ms}
&=&\tilde G_{\ms} ~\{ R_{\ms , \mf}(+)~ \tilde G_\mf ~R_{\mf , \ms}(+) \},
\label{12} \\
\tilde G_\mf~ \tg_{\ms} &=&\tg_{\ms} ~\{ R_{\ms , \mf}(+)~ \tilde G_\mf~
 R_{\mf,\ms}(+) \} ,\label{13} \\
 \tg^{-1}_{\ms}~\tilde G_\mf &=& \{ R_{\ms , \mf}(+)~ \tilde G_\mf ~R_{\mf ,
\ms}(+) \}
{}~\tg^{-1}_{\ms},\label{14}
\end{eqnarray}
where we use the curly bracket to bracket relevant quantities together
to make the meaning of equations clearler.
Associativity for the products of the fields are true  because
the $R$ matrix satisfies the   Yang-Baxter relations
 (which we leave as an excercise for the reader.)
Eqs.~(\ref{12}) to (\ref{14}) are the algebraic relations for $\tilde G$
parallel  to
those of
Eqs.~(\ref{7})  to (\ref{9}) for $\tilde j$.

 The basic elements of the quantum-group generators
$\{\te_i;i=3, ~{\rm and}~\pm\}$ are  related
to the components of the matrix
$\tilde G$ by
\begin{equation}
\tilde G \equiv
\left(  \begin{array}{cc}
          1 & 0 \\
          (1-q^2) ~\te_+ & 1  \end{array} \right)
\left(  \begin{array}{cc}
          q^{-\te_3} & 0 \\
          0 & q^{\te_3}      \end{array} \right)
\left(  \begin{array}{cc}
          1 & (q^{-1}-q) ~\te_- \\
          0 & 1              \end{array} \right)
,   \label{G}\end{equation}
where $\te_\pm$ and $q^{-{\te_3}}$ satisfy the standard quantum-groups
algebras \cite{dj,com3} which guarantee Eqs.~(\ref{12}) to (\ref{14}).

\smallskip

{\bf \it Semi-local Quantum-Group Generators}

Changing the integration range of Eq.~(\ref{11}) to a finite one, we obtain
the
semi-local
quantum-group generators

\begin{equation}
 \tilde G^\Delta(\bar y)\equiv  \vec P exp(
\int_{\bar y-\Delta}^{\bar y+\Delta}
%%%\int_{\scriptstyle-\infty\atop\scriptstyle \bar y-\Delta}^{\bar y+\Delta}
d \bar y'~ \tg^{-1}\partial_{\bar y'}\tg)
=\tg^{-1}(\bar y-\Delta)~\tg(\bar y +\Delta).
\label{15}\end{equation}
Again using the exchange algebras, Eqs.~(\ref{1}), (\ref{5a}), and
(\ref{5b}), we obtain
\begin{eqnarray}
\{ R^{-1}_{\mf , \ms}&&
\!\!\!\!\!\!\!\!\!\!\!\!\!\!
(\bar y_1-\bar y_2) ~\tm_\mf(\bar y_1) ~R_{\mf
,
\ms}(\bar y_1-\bar y_2+2\Delta) \}~
\tm_{\ms}(\bar y_2)  \nonumber\\
&=& \{ \tm_{\ms}(\bar y_2) ~R^{-1}_{\mf , \ms}(\bar y_1-\bar y_2-2\Delta)~
\tm_\mf(\bar y_1) \}~ R_{\mf ,
\ms}(\bar y_1-\bar y_2),
 \label{16} \\
 \tm_\mf (\bar y_1)~ \tg_{\ms}(\bar y_2) &=&\tg_{\ms}(\bar y_2) ~\{
R^{-1}_{\mf
, \ms}
(\bar y_1-\bar y_2-\Delta)
 ~\tm_\mf(\bar y_1)~ R_{\mf , \ms}(\bar y_1-\bar y_2+\Delta) \} ,\label{17}
\\
\tg^{-1}_{\ms}(\bar y_2)~\tm_\mf(\bar y_1)&=& \{ R^{-1}_{\mf , \ms}(\bar
y_1-\bar
y_2-\Delta)~ \tm_\mf(\bar y_1) ~R_{\mf ,
 \ms}(\bar y_1-\bar y_2+\Delta) \}
 ~\tg^{-1}_{\ms}(\bar y_2). \label{18}\end{eqnarray}

Next we express the semi-local generator in terms of its annihilation and
creation
parts following a procedure similar to that used in Ref.~\cite{fr},
\begin{equation}
\tm (\bar y) \equiv [G^{\Delta +}(\bar y)]^{-1} G^{\Delta
-}(\bar y)~~,\label{19}\end{equation}
\noindent where the operators $G^{\Delta\pm}(\bar y)$ satisfy the following
exchange
algebras that guarantee Eqs.~(\ref{16}) to (\ref{18}),
\begin{eqnarray}
R_{\mf ,\ms}(\bar y_1-\bar y_2)~ G^{\Delta \pm}_{\mf}(\bar y_1)~ G^{\Delta
+}_{\ms}(\bar y_2)
&=& G^{\Delta \pm}_{\ms}(\bar y_2)~G^{\Delta \pm}_{\mf}(\bar y_1)~ R_{\mf
,\ms}(\bar y_1-\bar y_2),\label{20}\\
R_{\mf ,\ms}(\bar y_1-\bar y_2+\Delta)~ G^{\Delta +}_{\mf}(\bar y_1)
{}~G^{\Delta -}_{\ms}(\bar y_2)
&=& G^{\Delta -}_{\ms}(\bar y_2)~ G^{\Delta +}_{\mf}(\bar y_1)~ R_{\mf
,\ms}(\bar y_1-\bar y_2-\Delta),\label{21}\\
g_{\mf}(\bar y)~G^{\Delta\pm}_{\ms}(\bar y_2) &=& G^{\Delta\pm}_{\ms}(\bar
y_2)~ g_{\mf}(x) ~R_{\mf ,\ms}(\bar y_1-\bar y_2 \pm \Delta/2).\label{22}
\end{eqnarray}

Notice that
\begin{equation}
\biggl[\sum\nolimits^{n=+\infty}_{n=-\infty} \tilde j(\bar
y+n\Delta),G^\Delta(\bar
y)\biggr]=0,\label{23}\end{equation}
which manifests what we call the sl$^{\Delta}(n)\otimes Uq^{1/\Delta}
[\hbox{sl}(n)]$
symmetry of the theory.  For
$\Delta\rightarrow \infty$, Eq.~(\ref{23}) becomes $[\tilde j(\bar y),\tilde
G]=0$, manifesting the $\widehat{\hbox{sl}(n)}\otimes Uq[\hbox{sl}(n)]$
symmetry
of the theory. For $\Delta\rightarrow 0$, Eq.~(\ref{23}) becomes
$[\tilde Q,\tilde j^q(\bar y)]=0$, where $\tilde Q\equiv
\int\limits^\infty_{-\infty}\tilde j(\bar y)d\bar y={\lim\atop{\Delta\to
0}}\sum^\infty_{n=-\infty}[\Delta \tilde j(\bar y+n \Delta)]$ and $\tilde
j^q(\bar y)$ is from the coefficient of the $\Delta$-term in the
expansion of the right-hand-side of Eq.~(\ref{15}), manifesting the
$\hbox{sl}(n)\otimes U^{\infty}_q [\hbox{sl}(n)]$ symmetry of the theory.

\smallskip

{\bf \it Quantum-Group Difference
Equation for Correlation Functions Defined in
\linebreak
the \hbox{ $U_q^{\Delta}[{\rm
sl}(2)]$-Vacuum} }

Using Eqs.~(\ref{19}) and (\ref{4}), we can obtain from Eq.~(\ref{15})
\begin{equation}
\tg(\bar y+\Delta) = \tg(\bar y-\Delta) \tm(\bar y)
=\tg(\bar y -\Delta)[\tilde G^{\Delta +}(\bar y)]^{-1}G^{\Delta
-}(\bar y).\label{24}\end{equation}

\noindent Since we are interested in the
vacuum expectation values of the products of the $\tilde g$ fields in the
$U_q^{\Delta}[\hbox{sl}(2)]$-vacuum state $\mid
0_q\rangle$ defined by
\begin{equation}
G^{\Delta -}(\bar y)\mid 0_q\rangle~=~\mid 0_q\rangle~,~~~{\rm
and}~~~\langle0_q\mid G^{\Delta +}(\bar y)~=~\langle0_q\mid~.\label{25}
\end{equation}
we want to move $\left[G^{\Delta +}\left(\bar y\right)\right]^{-1}$
to the left of $\tilde g(\bar y-\Delta)$ in Eq.~(\ref{24}). To achieve that
feat we
use
Eq.~(\ref{22}), many matrix relations, and finally  reach the goal
\begin{equation}
\tilde g(\bar y+\Delta)= \left(\left(\left(G^{\Delta +}(\bar
y)\right)^{-1}\right)^T
\Upsilon~\tg^T(\bar y-\Delta)\right)^T G^{\Delta -}(\bar y),\label{26}
\end{equation}
\noindent where the superscript $T$ means matrix transposition (the order
of the operators stays put);
$\Upsilon \equiv {q+q^{-1} \over q^2+q^{-2}}
\times diag(q,q^{-1})$ resulted from
\begin{equation}
\Upsilon_{\mf} = (Tr)_{\ms}\left(P_{\mf ,\ms}\left(\left(\left(R_{\mf
,\ms}(0)\right)^{T_\mf}\right)^{-1}\right)^{T_{\ms}}
\right)~,\label{27}\end{equation}
where the superscripts $T_\mf$ and $T_{\ms}$ indicate transpose of
matrices in
the tensor spaces $I$ and $II$ respectively.

Using Eq.~(\ref{25}) and Eq.~(\ref{26}), we obtain the quantum-group
difference equation
for the
correlation functions
\[
\langle 0_q\mid \tg_\mf (\bar y_1) \cdots \tg_L (\bar y_l+2\Delta) \cdots
\tg_N(\bar y_n)
\mid 0_q\rangle=\langle 0_q\mid \tg_\mf (\bar y_1) \cdots \tg_L (\bar y_l)
\cdots
\tg_N(\bar y_n) \mid
0_q\rangle\]
\begin{equation}
\times R_{L,L-1}(\bar y_l-\bar y_{l-1}) \cdots R_{L,I}(\bar y_l-\bar
y_1) \Upsilon_L
R_{L,N}(\bar y_l-\bar y_n+2\Delta) \cdots R_{L,L+1}(\bar y_l-\bar
y_{l+1}+2\Delta).\label{28}\end{equation}
For the special case of $``+2\Delta$'' being with $\bar y_n$ on the left
side of
Eq.~(\ref{28}), Eq.~(\ref{28})  simplifies to the following cyclic relation
\begin{equation}
\langle 0_q\mid \tg_\mf (\bar y_1) \cdots \tg_N(\bar y_n+2\Delta) \mid
0_q\rangle=\langle 0_q\mid \tg_N (\bar y_n) \tg_I (\bar y_1) \cdots
\tg_{N-1}(\bar
y_{n-1}) \mid  0_q\rangle\Upsilon_N~~.\label{29}\end{equation}
Similarly, difference equations for products involving the $\tilde g^{-1}$'s
can also
be obtained.

For the two point function, Eq.~(\ref{28}) becomes
\begin{equation}
\langle0_q\mid\tg_\mf(\bar
y_1)\tg_{\ms}(\byb+2\Delta)\mid0_q\rangle=\langle0_q\mid\tg_\mf(\bya)
\tg_{\ms}(y_2)\mid0_q\rangle R_{\ms
,\mf}(\byb-\bya)\Upsilon_{\ms}~~.\label{30}\end{equation}
\noindent Multiplying Eq.~(\ref{30}) from the right by ${\cal
P}^q_{j_{12}=0}$
and using the fact $\langle0_q\mid\tg_\mf\tg_{\ms}\mid0_q\rangle{\cal
P}^q_{j_{12}=0}$ $=\langle 0_q\mid\tg_\mf\tg_{\ms}\mid 0_q\rangle$, which
can be shown using the definition of $\mid 0_q\rangle$ given by
Eq.~(\ref{25}),
we obtain
\begin{equation}
\langle 0_q\mid \tg_\mf (\bar y_1) \tg_{\ms}(\bar y_2+2\Delta) \mid
0_q\rangle
=\langle 0_q\mid \tg_\mf (\bar y_1) \tg_{\ms}(\bar y_2)
\mid 0_q\rangle ~q^{-\Delta_0\varepsilon
(\bar y_1-\bar y_2)}
{q+q^{-1} \over q^2+q^{-2}},\label{31}\end{equation}
\noindent where the last factor on the right is from $R_{\mf
,\ms}(\bya-\byb)\Upsilon_{\ms}{\cal P}^q_{j_{12}=0}={\cal P}^q_{j_{12}=0}
q^{-{\Delta_0}\varepsilon(\bya-\byb)}(b/a)$ with $b/a\equiv
(q+q^{-1})/(q^2+q^{-2})=([2]_q)^2/[4]_q$ and the fact that
${\cal P}^q_{j_{12}=0}$ multiplying the vacuum expectation value becomes
unit.

We find the solution to Eq.~(\ref{31}). It can be written in the following
form
\pagebreak
\[
\langle 0_q\mid\tg_\mf(\bya)\tg_{\ms}(\byb)\mid 0_q\rangle\]
\hskip.25in
\begin{equation}
=A_0 Exp \biggl\{ -\biggl({\bya-\byb\over
2\Delta}\biggr)ln\biggl({q+q^{-1}\over
q^2+q^{-2}}\biggr)+\biggl[\biggl({\bya-\byb\over
2\Delta}\biggr)+2\sum^\infty_{n=1}\theta\biggl(-{\bya-\byb\over
2\Delta}-n\biggr) \biggr] ln(q^{\Delta_{0}})\biggr\},\label{32}\end{equation}
\noindent where $A_0$ is an arbitrary constant; $\theta (x)=0,{1\over
2},1$ for $x<0, x=0, x>0$, respectively. This expression for the
solution is continuous in the $\bya-\byb>0$ region.  For expressing
the solution in a function that is continuous in the $\bya-\byb<0$ region,
we replace $(\bya-\byb)$ by $-(\bya-\byb)$ and
$\sum^\infty_{n=1}$ by $\sum^\infty_{n=0}$ in the square bracket of
the above equations.

%\vfill

\smallskip

{\bf \it Concluding remarks}

The group-valued local quantum fields $\tilde g$ and
$\tilde g^{-1}$ and their exchange
algebras form the foundation of the quantum WZNW field theory.
Understanding the meaning of the spatial dependence of
the $R$ matrix of the exchange algebras is essential in our formulation of
the
theory. Being bimodule quantum fields
and having quantum-group structures are generic features of non-Abelian
group-valued local quantum fields. From these group-valued quantum fields,
the
content of the theory in Lie-algebra-valued fields  can  easily be derived.
The
other way around is much harder.  The clear exposition of the bimodule
properties of group-valued fields in our formulation leads us to the explicit
construction of  the quantum-group generators,  the
$U_q^{\Delta}[\hbox{sl}(n)]$-vacuum, and the derivation of the quantum-group
difference equations for the correlation functions defined in the
$U_q^{\Delta}[\hbox{sl}(n)]$-vacuum. This way of understanding the quantum
WZNW
theory has served us well in developing the quantum
Self-dual Yang-Mills theory, which is a  quantum field theory with
interactions in
four dimensions.

 Quantum states of non-abelian
group-valued local
fields may well exist in nature. It
is important to look for them. If they do exist, we would like to call them
bimodulons \cite{c2}.

\bigskip

\medskip
\centerline{\bf Acknowledgment}

We would like to thank John H. Schwarz for helpful comments.

This work is supported in part by the U.S. Department of Energy (DOE).

\vfill


\begin{thebibliography}{99}
\bibitem{com1}
The use of boson quantum fields to represent $\tilde g$ introduces many
unnecessary
complications which hinder the full development of the theory.
\bibitem{com2}
For the sl$(n)$ cases all one needs to do is to replace the two
projection matrices in Eq.~(\ref{2}), ${\cal P}^q_{j_{12}=1}$ by the
symmetric
projection matrix ${\cal P}^q_s$ and ${\cal P}^q_{j_{12}=0}$ by the
antisymmetric projection matrix ${\cal P}^q_a$ of sl$(n)$, and $\Delta_1$ by
$({1\over n}-1)$ and $\Delta_0$ by $({1\over n}+1)$.
\bibitem{com3}
In Ref.~\cite{cy3} we discussed properties of such quantum-group generators
in the
SDYM
quantum field theory.  Now we have also constructed such generators for
the SDYM case, Ref.~\cite{cy4}.
\end{thebibliography}
\end{document}